\documentclass[aip,jcp,reprint,floatfix]{revtex4-1}
\usepackage{graphics,graphicx,amsfonts,amsmath,amsbsy,amssymb,color}
\usepackage{bm}
\usepackage{epic}
\usepackage{mciteplus}
\usepackage{subfigure}
\usepackage{psfrag}
\usepackage{multirow}
\usepackage{vector}  % Allows "\bvec{}" and "\buvec{}" for "blackboard" style bold vectors in maths

\def \beq {\begin{eqnarray}}
\def \eeq {\end{eqnarray}}
\def \Schrodinger {{Schr\"{o}dinger }}

\def \bfj {{\bf j}}
\def \bfi {{\bf i}}

\newcommand{\bra}{\ensuremath{\langle}}
\newcommand{\ket}{\ensuremath{\rangle}}

\newcommand{\rff}[1]{{Eq.~\eqref{#1}}}

\begin{document}
%\title{An Unbiased Approach to the Sampling of Properties in Full Configuration Interaction Quantum Monte Carlo via Reduced Density Matrices and Replica Samples}
\title{Unbiased Reduced Density Matrices and Electronic Properties from Full Configuration Interaction Quantum Monte Carlo}
\author{Catherine~Overy}
\affiliation{University of Cambridge, Chemistry Department, Lensfield Road, Cambridge CB2 1EW, United Kingdom}
\author{George~H.~Booth}
\email{george.booth24@gmail.com}
\affiliation{University of Cambridge, Chemistry Department, Lensfield Road, Cambridge CB2 1EW, United Kingdom}
\affiliation{King's College London, Theory and Simulation of Condensed Matter, The Strand, London WC2R 2LS, United Kingdom}
\affiliation{Thomas Young Centre, University College London, 17 Gordon Street, London WC1H 0AH, United Kingdom}
\author{N.~S.~Blunt}
\affiliation{University of Cambridge, Chemistry Department, Lensfield Road, Cambridge CB2 1EW, United Kingdom}
\author{James~J.~Shepherd}
\affiliation{University of Cambridge, Chemistry Department, Lensfield Road, Cambridge CB2 1EW, United Kingdom}
\author{Deidre~Cleland}
\affiliation{University of Cambridge, Chemistry Department, Lensfield Road, Cambridge CB2 1EW, United Kingdom}
\affiliation{CSIRO Virtual Nanoscience Laboratory, 343 Royal Parade, Parkville, Victoria 3052, Australia}
\author{Ali~Alavi}  
\email{asa10@cam.ac.uk}
\affiliation{University of Cambridge, Chemistry Department, Lensfield Road, Cambridge CB2 1EW, United Kingdom}
\affiliation{Max Planck Institute for Solid State Research, Heisenbergstr. 1, D-70569, Stuttgart, Germany}

\begin{abstract}
Properties that are necessarily formulated within pure (symmetric) expectation values are difficult to calculate for projector quantum Monte Carlo approaches, 
but are critical in order to compute many of the important observable properties of electronic systems. Here, we investigate an approach for the sampling of unbiased 
reduced density matrices within the Full Configuration Interaction Quantum Monte Carlo dynamic, which requires only small computational overheads. This is 
achieved via an independent replica population of walkers in the dynamic, sampled alongside the original population. The resulting reduced density matrices are 
free from systematic error (beyond those present via constraints on the dynamic itself), and can be used to compute a variety of expectation values and properties, 
with rapid convergence to an exact limit. A quasi-variational energy estimate derived from these density matrices is proposed as an accurate alternative to the 
projected estimator for multiconfigurational wavefunctions, while its variational property could potentially lend itself to accurate extrapolation approaches in larger 
systems.
\end{abstract}

\date{\today}
\maketitle

%To check in the code:
	% HPHF and RDMs working
	% Energy estimators and error analysis performed separately, but then combined for you in the output (currently a formatting error)
	% Popsfiles, semi-stochastic and all 'reals' cutoffs working as expected
	% Working with new excitation generators? (should be fine??)
	% Both contributions from the sampling are considered, from replica 1 and 2.
	% For the replicas, the running totals of the diagonals should only be accumulated if both replicas are occupied, and the summing in of the 
	%	diagonal contributions to the RDMs should only be done when one of the blocks becomes unoccupied.
\section{Introduction}

The extraction of expectation values and properties of quantum systems whose operator does not commute with the Hamiltonian has been a significant 
hurdle in the widespread adoption of projector Quantum Monte Carlo methods for many years\cite{Rothstein2013}. 
These include important physical properties such as the dipole or higher electrical moments, as well as particle distribution functions, 
forces and higher derivatives of nuclei position, static correlation functions, quantum entropy estimators and various order parameters in the condensed phase. 
As such, their reliable and unbiased computation within the projector Quantum Monte Carlo (QMC) framework is an active and important research 
area\cite{Rothstein1988,Reynolds1990,Lester1991,Lester1992,Langfelder,Reptation,HFSampling,HFSampling2}. 
Within continuum Diffusion Monte Carlo and its variants, there exist techniques to unbias these quantities such as `forward-walking' and its 
variants\cite{Reynolds1990,Lester1991,Lester1992,Langfelder}, Reptation Quantum Monte Carlo\cite{Reptation}, 
and Hellmann--Feynman sampling\cite{HFSampling,HFSampling2}, all of which have drawbacks regarding the additional computational effort they require. 

Recently, the Full Configuration Interaction Quantum
Monte Carlo method (FCIQMC) has been introduced as a projector QMC technique formulated within the second quantized algebra of arbitrary 
(antisymmetrized) Hilbert spaces. 
This method relies on the fact that, as opposed to continuum approaches, the space is finite (although exponentially large) and therefore can employ annihilation (cancellation) 
processes between walkers of different signs in order to overcome the Fermion sign problem\cite{BTA2009,BA2010,Spencer2012}. Additional systematically 
improvable approximations have also been 
introduced to the method to allow for a large reduction in computational effort, while nevertheless converging in the limit of large sampling to exact results within 
the basis employed\cite{CBA2010,BoothC2,CBOA2012}. Furthermore, there is significant empirical evidence that the computational effort required for the 
sampling of many systems scales 
sub-linearly with the size of the Hilbert space as systems grow in size and complexity\cite{BoothC2,CBA2011,Shepherd2014}. This has allowed for near-exact results to be obtained 
within systematically 
improvable random error bars with substantially less computational effort than an iterative diagonalization approach to the problem. This has allowed for large molecular\cite{Filippi2012}, 
solid state\cite{BGKA2013} and model Hamiltonians\cite{Shepherd2012_1,Shepherd2012_2,Spencer2012,Clark2012} to be investigated with the technique.

However, studies to date have primarily focused on properties of systems related to total energy differences, such as ionization potentials, electron affinities, potential
energy surfaces and equations of state, all of which can be computed in an unbiased fashion through calculation of a {\em projected} energy expression. This relies on the fact
that for an eigenstate $|\Psi \ket$, with energy $E$, this energy can be reproduced from
\begin{equation}
    E = \bra E_P \ket = \frac{\bra \psi_T|{\hat H}|\Psi \ket}{\bra \psi_T|\Psi \ket},	\label{eqn:ProjE}
\end{equation}
for any function $|\psi_T \ket$ with a non-zero overlap with $|\Psi \ket$.
However, to extend the scope of
the method, it is necessary to be able to extract unbiased properties from the sampled wavefunctions, for operators which do not commute with the Hamiltonian, such as those
which are a function of position. In these cases, Eq.~\ref{eqn:ProjE} will not be sufficient if the operator of interest
has a different set of eigenstates\cite{Lester1991}. 

Formally, these properties can be derived from the presence of a perturbing operator to the Hamiltonian, ${\hat A}$, as
\begin{equation}
\hat{H^{\prime}} = \hat{H} + \lambda \hat{A}	,
\end{equation}
where $\lambda$ is the strength of the perturbation. The relevant expectation value associated with ${\hat A}$ can then be obtained by calculating the analytical 
(or numerical) gradient of the energy with respect to the perturbation strength, at zero perturbation~\cite{Pulay2014},
\begin{equation}
\bra A \ket = \frac{\partial E^{\prime}}{\partial \lambda} \Big|_{\lambda=0}		.
\end{equation}
In the case of stationary wavefunctions, which is the case (on average) in FCIQMC, the Hellmann--Feynman theorem\cite{HellFeyn} reduces this to the calculation of a 
{\em pure} expectation value of the operator, ${\hat A}$, as
\begin{equation}
\bra A \ket = \frac{\bra \Psi | {\hat A} | \Psi \ket}{\bra \Psi | \Psi \ket} .		\label{eqn:PureEx}
\end{equation}
Equivalently, this can be written as the trace of ${\hat A}$ with the appropriate rank reduced density matrix for the number of coupled particles in ${\hat A}$.
In continuum approaches, the sampled distribution consists of the desired wavefunction multiplied by a trial wavefunction\cite{Foulkes2001}, whereas in FCIQMC, 
a single (pure) wavefunction distribution is sampled
\cite{footnote1}.
%\cite{We note that there has been recent work to also sample from a mixed distribution within an FCIQMC framework\cite{Clark2012}. However, 
%as mentioned, this difference is not fundamental to the inherent bias or approach presented in this paper, which can also be employed within mixed distribution sampling of FCIQMC.}. 
However, this does not remove an inherent bias from these expectation values, which fundamentally arises from the difficulty in evaluating quadratic functionals 
of stochastically sampled wavefunctions\cite{Rothstein1988}, as found in Eq.~\ref{eqn:PureEx}. Strictly, a similar bias arises even in naive accumulations of the projected energy
estimator of \rff{eqn:ProjE}. This is because the ratio of the sampled wavefunction distributions has a finite covariance which needs to be properly accounted for when calculating
the quantity\cite{Spencer12,Lee11}.

In this paper, we describe the source of this bias for quadratic functionals, before demonstrating a way to sample one- and two-body reduced density matrices (RDMs)
of the sampled wavefunction within the FCIQMC dynamic in an unbiased fashion, with only small additional computational overheads. This allows for calculation of 
{\em pure} expectation values and quantities of interest via a trace of the resultant $N$-representable reduced density matrices with the appropriate 
operator\cite{Coleman2007,Mazziotti2007,Blum2012}.
The accumulation of the one- and two-body RDMs therefore allows for access to one- and two-body expectation values of the wavefunction, assuming that the appropriate operator
can be projected into the space. 

Indeed, the energy is also a two-body expectation value (of the Hamiltonian), and therefore an alternative, variational
estimate of the energy is also accessible from the two-body RDM. There may be significant advantages to an accurate computation of this quantity, despite the availability of 
the projected estimate (Eq. 1), since the pure estimate does not rely on a good quality trial wavefunction in order to obtain a large overlap with the true wavefunction 
and minimize stochastic fluctuations of the energy. This is likely to be important in multiconfigurational systems with significant static correlation, whilst the variationality of the estimator
may also offer a more rigorous extrapolation procedure for systematic errors compared to the non-variational projected estimate. Furthermore, as a quadratic functional of the 
wavefunction, errors in the energy given by the pure expectation value of Eq.~\ref{eqn:PureEx} will scale quadratically with the error in the wavefunction, rather than linearly in the case 
of Eq.~\ref{eqn:ProjE}, again potentially leading to an improved estimator.

Section~\ref{sec:FCIQMCRecap} will briefly recap the stochastic FCIQMC algorithm, before detailing an adaptation to allow for non-integer walker weights to reduce the
random errorbars in the approach. Section~\ref{sec:SamplingRDMs} will then detail a naive approach to sampling density matrices, which was implemented in 
Ref.~\onlinecite{Booth2012},
but which was known to be biased, the origin of which is pinpointed and discussed. An updated approach is then formulated and presented in section~\ref{sec:newRDMmethod_algo},
which rigorously removes this bias, and the convergence and unbiased nature of the resulting properties are demonstrated. The computational overhead of the
approach is analysed in detail, and strategies to minimize its impact without introducing further approximations illustrated. Finally, rigorous benchmarking of a range of
properties are shown, including the quasi-variational energy estimate, all of which show encouraging accuracy and convergence properties with both imaginary time, and walker number.

\section{FCIQMC recapitulation}	\label{sec:FCIQMCRecap}
%%%%%%%%%%% FCIQMC Recap %%%%%%%%%%% 

Full Configuration Interaction Quantum Monte Carlo can be considered as a stochastic minimization of the energy with respect to a sampled full configuration interaction 
wavefunction expansion. This wavefunction is a simple linear combination of all Slater determinants that can be constructed from distributing the available electrons within 
the (orthonormalized) single-particle orbitals spanning the space, as
\begin{equation}
|\Psi \ket = \sum_{\bfi} C_{\bfi} | D_{\bfi} \ket	,	\label{eqn:FCI}
\end{equation}
where $|D_{\bfi} \ket$ represents a Slater determinant, labeled by the orbital occupation string $\bfi$.
The linear coefficients of this expansion are the objects that are stochastically sampled, such that the average sampled determinant weights
$\bra n_{\bfi} \ket$ are proportional to $C_{\bfi}$.
Their optimization to the variational ground state of the ansatz is simulated via a stochastic, iterative application of the equations:
\begin{equation}
\Delta n_{\bfi}(\beta+\tau) = -\tau \left[ \sum_{\bfj \ne \bfi} H_{\bfi \bfj} n_{\bfj}(\beta) \right]- \tau (H_{\bfi \bfi} - E_S) n_{\bfi}(\beta)		\label{eqn:master}
\end{equation}
where $\Delta n_{\bfi}(\beta)$ represents the change in `walker' population/weight on determinant $|D_{\bfi} \ket$ in the time step $\beta \rightarrow \beta+\tau$.
This leads to population dynamics of a set of walkers which occupy determinants connected to each other in this many-electron Hilbert space. 
This dynamic consists of a set of stochastically realized processes. 

The first is a `spawning' step, which is performed for each occupied determinant, and a number of times proportional to the walker weight at that determinant ($n_{\bfi}$). A 
single or double excitation is randomly chosen, with normalized probability $p_{\textrm{gen}}(\textbf{j}|\textbf{i})$ for the excitation from $|D_{\textbf{i}}\ket$ to
$|D_{\textbf{j}}\ket$. The walker amplitude on $|D_{\textbf{j}}\ket$ is then augmented with a signed probability given by
\begin{equation}
p_{\textrm{spawn}} = - \frac{\tau H_{\bfi \bfj}}{p_{\textrm{gen}}(\textbf{j}|\textbf{i})}	.	\label{eqn:accprob}
\end{equation}
Finally, a `death' step is performed, by which the amplitude on each determinant, $|D_{\bfi}\ket$,  is (generally) reduced with probability $\tau (H_{\bfi \bfi} - E_S) n_{\bfi}$.
Taken together, these two steps simulate the dynamic in Eq.~\ref{eqn:master}. However, an additional `annihilation' step is essential in order to overcome an
exponential increase in noise and other features associated with the Fermion sign problem\cite{Spencer2012,KolodrubetzSpencerClarkFoulkes2013}. In this step, 
walkers of opposite signs on the same determinant are removed from the simulation.

%These consist of spawning steps (to determinants connected through 
%the off-diagonal elements of the Hamiltonian operator), death steps (which augment the
%walker populations on the individual determinants according to their energy), and annihilation steps (where walkers of opposite signs on the same determinant
%are removed from the simulation, in order to overcome an exponential increase of noise and other features of the Fermion sign problem\cite{}).

As mentioned in the introduction, the energy of the wavefunction can be extracted from a projected estimator (Eq.~\ref{eqn:ProjE}), which in this paper is simply 
obtained from a projection onto the Hartree--Fock determinant ($|\psi_T \ket=|D_{\textrm HF} \ket$). In addition, the value of $E_S$ is varied throughout the simulation in
order to maintain a constant, desired weight of walkers. At convergence, this value should fluctuate about the energy of the system, providing an alternative estimator
for the energy based on the total growth rate of all the walkers in the system.
More details on the specific implementation of these steps, and the
derivation of this dynamic from the imaginary-time \Schrodinger equation can be found in 
Refs.~\onlinecite{BTA2009,BSA2014,PetruzieloHolmesChanglaniNightingaleUmrigar2012}. Furthermore, the systematically improvable {\em initiator}
approximation is used exclusively throughout this work (sometimes denoted $i$-FCIQMC to distinguish it from the full method). 
This involves a dynamically truncated Hamiltonian operator, where spawning events to unoccupied determinants
are constrained to be only allowed if they come from a determinant with a walker weight greater than $n_{\textrm add}$. This approximation can be systematically improved
as the number of walkers increases, as increasing numbers of determinants fulfil the criteria, and the sampled Hamiltonian therefore approaches the exact Hamiltonian.
More details and benchmarking of this approximation can be found in Refs.~\onlinecite{CBA2011,BoothC2}.

% with reals
\subsection{Non-integer walker weights}	\label{sec:reals}

Recent advances in the methodology were introduced by Petruzielo {\em et. al.}\cite{PetruzieloHolmesChanglaniNightingaleUmrigar2012}, 
where among other things, non-integer walker weights were introduced. This non-integer 
extension to FCIQMC requires a realization of Eq.~\ref{eqn:master} which allows walkers with non-integer weights to be spawned. This
is done by applying the spawning and death processes continuously rather than discretely. By `continuous' in this context, we mean that weights are assigned in a spawning/death
process in a continuous framework, without the additional stochastic process to convert the
resultant weight into an integer value for $\Delta n_{\bfi}$. Applying both of these steps continuously removes the need for much of the
random number generation in the code, and reduces the instantaneous fluctuations on any given determinant by allowing modification of its population
by small fractions of a walker, rather than by whole walkers at a time. It should also be noted that when determining the number of {\em attempted} spawning events 
from a determinant with a non-integer number of walkers, the fractional part of the amplitude is used as a stochastic test for whether to perform an 
additional spawning step on top of $\lfloor n_{\textbf{i}} \rfloor$, to ensure the overall number of spawning attempts is still proportional to $n_{\bfi}$, where $\lfloor \quad \rfloor$ and
$\lceil \quad \rceil$ represent rounding down or up to the next integer value of a real number respectively.

However, with continuous spawning events automatically accepted (unless $|H_{\bfi \bfj}|$ is exactly zero), the
number of successful spawning events is substantially increased. This requires more memory to store the increase in spawned walkers, but also slows
the simulation down, as each spawning event must be put through some combination of communication and annihilation steps in order to transfer this
information. It is therefore necessary to have some level of stochastic compression of the low-weighted walkers, to avoid storage costs for walkers which would 
quickly span much of the space.

In this work, the overhead of dealing with the additional low weighted walkers is controlled by two parameters. First, a subspace over which the non-integer walker weights are allowed
is imposed. This is most simply done via a restriction of non-integer weights to only be allowed within a subspace defined by a given number of particle-hole excitations from the
reference determinant (generally Hartree--Fock), determined by the cutoff parameter $\chi$. 
This greatly minimizes the fluctuations in $E_P$, since any value of $\chi \ge 2$ will ensure that all determinants which contribute to $\langle E_P \rangle$ (with the
Hartree--Fock trial wavefunction) are included in the non-integer space, which is finely resolved. Any spawning event out of this space requires an additional stochastic step
to discretize the walker amplitude to an integer quantity.

In addition, a minimum occupation threshold ($N_{\textrm{occ}}$) is defined, in order to ensure that low-weighted determinants do not proliferate, as was initially 
introduced in Ref.~\onlinecite{PetruzieloHolmesChanglaniNightingaleUmrigar2012}. After all annihilation events are complete, the instantaneous population of each 
determinant, $n_{\bfi}$, is assessed in comparison to this threshold. 
If $n_{\textbf{i}} < N_{\textrm{occ}}$, its population is rediscretised to either $N_{\textrm{occ}}$ (with probability $\frac{n_{\textbf{i}}}{N_{\textrm{occ}}}$) or 
$0$ (with probability $1 - \frac{n_{\textbf{i}}}{N_{\textrm{occ}}}$). Larger values of $N_{\textrm{occ}}$ will therefore reduce the occupied space of determinants 
by depopulating more determinants with $n_{\textbf{i}} < N_{\textrm{occ}}$.
Of course, values in excess of $1$ correspond to a \emph{coarser} representation of the wavefunction than the original integer algorithm in certain regions of the space, 
making such a choice inadvisable.
Conversely, setting $N_{\textrm{occ}}=0$ has no effect at all and returns us to the untenable scenario where a very 
large number of instantaneously occupied determinants with low weight must be simultaneously stored.

An alternative strategy is to restrict the total number of successful spawning events, which is known to be much higher in 
the non-integer algorithm. Many of these events may be propagating only small fractions of a walker at a time.
For spawning events in particular, the cost of treating these low weight progeny in terms of storage and communication, can be substantial compared to their negligible 
impact on the progress of the simulation. The death events do not bear the same cost, as the populations are directly updated in the main list of determinants.
Therefore, a minimum threshold value, $\kappa$, can be introduced for a continuous spawning event, with the advantage over $N_{\textrm{occ}}$ that additional communication
costs are avoided (these strategies can of course be used together).
If $p_{\textrm{spawn}}$ (as given in Eq.~\ref{eqn:accprob}) is \emph{greater} than $\kappa$, $p_{\textrm{spawn}}$ walkers are spawned as a continuous event. 
If $p_{\textrm{spawn}}$ is \emph{less} than $\kappa$ (i.e. a low weight event)  $\kappa$ walkers are spawned with probability $\frac{p_{\textrm{spawn}}}{\kappa}$, otherwise $0$ walkers are spawned.
By doing this, the disproportionate cost of treating progeny smaller than $\kappa$ is saved.

It was found that the description of at least some parts of the space in this continuous fashion, allowing for finer resolution of these wavefunction amplitudes, was
universally beneficial, with efficiency gains of up to 120 times over full use of integer walker weights. This was due to the reduction in stochastic noise, 
although the precise efficiency gain
was significantly system-dependent. The change was also not found to have any negative effect on equilibration or serial correlation times. The gain in efficiency 
was found to be fairly consistent across values for $\chi$, $N_{\textrm{occ}}$ and $\kappa$, however, an approach with $\chi=4$ (for a CISDTQ non-integer subspace) 
and a minimum occupation of 1 walker ($N_{\textrm{occ}}=1$) was found to be advisable due to its more modest memory demands compared to a more complete 
non-integer walker space. This recommended approach typically resulted in 50\% increased computational effort, and approximately twice as much memory than the 
integer method, which is more than offset by the gains in efficiency. These parameters will be used throughout this work unless otherwise specified, with $\kappa=0$. A more detailed 
analysis of this efficiency gain can be found in Ref.~\onlinecite{OveryThesis2014}.

\section{Stochastic Sampling of Reduced Density Matrices}	\label{sec:SamplingRDMs}
%%%%%%%%%%% RDMs Overview %%%%%%%%%%%

The second-order reduced density matrix (2-RDM) can be expressed in second quantization within a basis of spin orbitals as a rank-4 tensor, $\bf{\Gamma}$,
\begin{equation}
\Gamma_{pq,rs} = \bra \Psi | {a}_p^{\dagger} {a}_q^{\dagger} {a}_s {a}_r | \Psi \ket \hspace{1cm} p > q , \hspace{3mm} r > s\label{eqn:compact2RDM}
\end{equation}
where ${a}_p^{\dagger}$ and ${a}_p$ are creation and annihilation operators respectively~\cite{Helgie}.
As the $N$-electron Hamiltonian is constructed only from 1- and 2-body operators, the 2-RDM representation contains sufficient information to exactly reproduce the total energy, as
\begin{equation}
    E_{\textrm{RDM}} = \sum_{pq} h_{pq} \gamma_{pq} + \sum_{p>q,r>s} \Gamma_{pq,rs} \bra pq || rs \ket + h_{\textrm{nuc}} \label{eqn:ERDM}
\end{equation}
where the 1-RDM is represented as $\gamma_{pq}=\bra \Psi | {a}_p^{\dagger} {a}_q| \Psi \ket$.
In addition to accessing the total energy, the 2-RDM can also be used to compute other 1- and 2-body static properties of the system, provided that the appropriate operator
can be projected into the space.
This is achieved by computing the trace of the relevant 1- or 2-electron operator $\hat{Q}$ with the RDM
\begin{equation}
    \bra Q \ket = \textrm{Tr} [{\bf \Gamma}\hat{Q}] 
\end{equation}
where $\hat{Q}$ is expressed in the same one-particle basis as $\bf{\Gamma}$ (Ref.~\onlinecite{Blum2012}).
As such, the reduced density matrices are a powerful construct, allowing the calculation of a variety of {\em pure} expectation values with a highly compact 
representation of the relevant information in the $N$-electron wavefunction, without introducing any approximation into the total energy or other observable 
quantities\cite{Coleman2007,Mazziotti2007,Blum2012}.
The reduced density matrices also form the basis of various measures of quantum entanglement~\cite{PopkovSalerno2012}, can be used to compute multireference 
explicit correlation corrections to reduce finite basis error\cite{Kong:CR112-75,Booth2012,Torheyden:JCP131-171103}, as well as provide the information for self-consistent
optimisations of orbital spaces, as required for the complete active space self-consistent field (CASSCF) approach\cite{KnowlesCASSCF,DMRGCASSCF}, and the density matrix embedding theory (DMET)\cite{KniziaChan2012,KniziaChan2013,Chen2014}.

For a FCI wavefunction expansion in a determinant basis with coefficients $C_{\textbf{i}}$, the 2-RDM can be written as
\begin{equation}
    \Gamma_{pq,rs} = \sum_{\textbf{ij}} C_{\textbf{i}} C_{\textbf{j}} \bra D_{\textbf{i}} | {a}_p^{\dagger} {a}_q^{\dagger} {a}_s {a}_r | D_{\textbf{j}} \ket \label{eqn:2RDMTerms}
\end{equation}
where the matrix element is non-zero only when $D_{\textbf{j}}$ is connected to $D_{\textbf{i}}$ by an $rs \rightarrow pq$ excitation, giving a value of $\pm 1$ (repeated
indices are also allowed). Diagonal terms in the RDMs can therefore be found as
\begin{equation}
    \Gamma_{pq,pq} = \sum_{\{p,q\} \hspace{1pt} \in \hspace{1pt} \textbf{i}} (C_{\textbf{i}})^2. \label{eqn:2RDMDiagTerms}
\end{equation}
Therefore, within the FCIQMC dynamic, each determinant, $D_{\textbf{i}}$, contributes to the 2-RDM via $C_{\textbf{i}}^2$ summed into each of the $\frac{N(N-1)}{2}$ elements 
where orbitals $p$ and $q$ are occupied in $D_{\textbf{i}}$, to generate the diagonal elements of the 2-RDM in a relatively simple fashion.

However, the explicit generation of all relevant determinant pairs in Eq.~\ref{eqn:2RDMTerms} for the off-diagonal contributions is
prohibitively expensive ($\mathcal{O}[N^2M^2]$ per occupied determinant), and so we seek an algorithm which avoids this naive approach. This was considered in
Ref.~\onlinecite{Booth2012}. As the 2-RDM involves contributions from pairs of determinants that are (at most) double excitations of one another, the existing 
spawning events that occur throughout the simulation to propagate the walker distribution in imaginary time already generate the required excitations, through the application of the
Hamiltonian operator. 

For each walker on determinant $D_{\textbf{i}}$, the spawning step generates a target determinant, $D_{\textbf{j}}$, which is always a single or double excitation of $D_{\textbf{i}}$.
If the spawning attempt is successful, the child walker is communicated to the parallel process that hosts $D_{\textbf{j}}$, allowing its weight to be updated.
The stochastic RDM method exploits this existing computational effort and communication to sample the RDM off-diagonal elements.
To calculate the contribution to the unnormalized RDM from this determinant pair, the signed amplitude information of both determinants is required, which 
can be achieved by passing 
a small amount of additional information through the existing communication step with newly spawned walkers (an amplitude and the identity of the parent 
determinant, $D_{\textbf{i}}$).
After the communication step, the identity and signed weight of both 
determinants are then available on the same process.
The identity of the orbitals that differ between them can then be easily calculated, and the contribution added to the relevant off-diagonal RDM element(s).
Therefore, contributions to the off-diagonal RDM terms are only included when there is a successful spawning attempt between two determinants. 

As the RDM contribution corresponding to $C_{\textbf{i}} C_{\textbf{j}}$ is only added in when there is a successful spawning event between the determinants, it 
must be rescaled to take into account the probability of this event occurring. The contribution added into these terms will therefore take the form
\begin{equation}
	 \frac{C_{\textrm{i}} C_{\textrm{j}}} {p_c(n_{\bfi})(\textbf{j}|\textbf{i})}
%    \frac{{\bra n_{\textbf{i}} \ket}_{\tau_{\textrm{occ}}} {\bra n_{\textbf{j}} \ket}_{\tau_{\textrm{occ}}}}{p_c(\textbf{j}|\textbf{i})}
\end{equation}
where $p_c(n_{\bfi})(\textbf{j}|\textbf{i})$ is the normalized probability of spawning at least one child (of any weight) onto $D_{\textbf{j}}$ from $D_{\textbf{i}}$ during the current iteration.
This depends on factors such as the number of walkers on $D_{\textbf{i}}$, and whether it is a stochastic or continuous spawning event. This can be calculated for non-integer weighted determinants as
\begin{equation}
p_c(n_{\bfi})(\textbf{j}|\bfi) = 1 - (\lceil n_{\textbf{i}} \rceil - n_{\textbf{i}}) \lambda^{\lfloor n_{\textbf{i}}\rfloor} - (n_{\bfi} - \lfloor n_{\textbf{i}}\rfloor) \lambda^{\lceil n_{\textbf{i}}\rceil} 
\end{equation}
and for determinants populated by integer numbers of walkers as
\begin{equation}
    p_c(n_{\bfi})(\textbf{j}|\bfi) = 1 - \lambda^{n_{\bfi}} 
\end{equation}
where $n_{\textbf{i}}$ denotes the instantaneous walker weight on determinant $D_{\textbf{i}}$ for the iteration in consideration, and $\lambda$ is the probability of {\em not}
spawning a walker (of any weight) from $D_{\bfi}$ to $D_{\bfj}$ in a single attempt. This value varies depending on what `type' of spawning event is being attempted. Specifically,
\begin{align}
\lambda_{\textrm{stoch.}} &= 1-\min(\tau |H_{\bfj \bfi}|,p_{\textrm{gen}}(\bfj|\bfi))	\label{eqn:StochUnbias} \\
\lambda_{\textrm{cont.}}   &= 1- p_{\textrm{gen}}(\textbf{j}|\textbf{i})	\label{eqn:ContUnbias}	\\
\lambda_{\textrm{spawn cutoff}} &= 1 - \frac{\tau |H_{\bfj \bfi}|}{\kappa}	\qquad \textrm{if} \quad p_{\textrm{spawn}} < \kappa	\label{eqn:SpawnCutUnbias}
\end{align}
where $\lambda_{\textrm{stoch.}}$ (Eq.~\ref{eqn:StochUnbias}) is to be used in the case of a stochastically realized integer spawning event, $\lambda_{\textrm{cont.}}$
(Eq.~\ref{eqn:ContUnbias}) is to be used for a continuous spawning event, and $\lambda_{\textrm{spawn cutoff}}$ (Eq.~\ref{eqn:SpawnCutUnbias}) is to be used if the 
continuous spawning truncation 
parameter, $\kappa$, is being used (see section~\ref{sec:reals}) and $p_{\textrm spawn} < \kappa$, otherwise $\lambda_{\textrm{cont.}}$ should be used. It should be noted that
the unbiasing factor $p_c(n_{\bfi})(\textbf{j}|\bfi)$ is not a constant, and is an implicit function of $\beta$ since it depends on the instantaneous walker weight $n_{\bfi}$,
for the iteration under consideration.
%\begin{align}
%    p_c(\beta)(\textbf{j}|\textbf{i})_{\textrm{stoch.}} &= 1 \hspace{-1pt} - \hspace{-1pt} (\lceil n_{\textbf{i}} \rceil \hspace{-1pt} - \hspace{-1pt} n_{\textbf{i}}) \Big [ 1\hspace{-1pt}-\hspace{-1pt}\tau |H_{\textbf{ji}}| \Big ]^{\lfloor n_{\textbf{i}}\rfloor} \hspace{-9pt}	\nonumber \\
%    & - (n_{\textbf{i}}\hspace{-1pt}-\hspace{-1pt}\lfloor n_{\textbf{i}}\rfloor) \Big [1\hspace{-1pt}-\hspace{-1pt} \tau |H_{\textbf{ji}}| \Big ]^{\lceil n_{\textbf{i}}\rceil} \hspace{-10pt} \label{eqn:pc_stoch} \\
%     p_c(\beta)(\textbf{j}|\textbf{i})_{\textrm{cont.}} &= 1 \hspace{-1pt}- \hspace{-1pt}(\lceil n_{\textbf{i}}\rceil \hspace{-1pt} - \hspace{-1pt} n_{\textbf{i}}) \Big [1 \hspace{-1pt} -\hspace{-1pt} p_{\textrm{gen}}(\textbf{j}|\textbf{i})\Big ]^{\lfloor n_{\textbf{i}}\rfloor} \hspace{-13pt} \nonumber	\\
%     & - (n_{\textbf{i}} \hspace{-1pt} - \hspace{-1pt} \lfloor n_{\textbf{i}}\rfloor) \Big [1\hspace{-1pt} - \hspace{-1pt} p_{\textrm{gen}}(\textbf{j}|\textbf{i}) \Big ]^{\lceil n_{\textbf{i}}\rceil} \hspace{-10pt} \label{eqn:pc_cont}
%\end{align}
%for stochastic and continuous spawning events respectively, where $n_{\textbf{i}}$ denotes the instantaneous walker weight on determinant $D_{\textbf{i}}$ for the 
%iteration in consideration.

If multiple successful spawning events are registered between a specific \{\bfi, \bfj\} pair in a single iteration, then it is necessary to only consider this to be a single contribution to the density matrix. In principle, if a pair of determinants contributes to the off-diagonal RDM element $\Gamma_{pq,rs}$, it will also contribute to the element $\Gamma_{rs,pq}$.
However, in practice, we choose to only update one of these terms for a spawning event, with the other sampled by events $D_{\textbf{j}} \rightarrow D_{\textbf{i}}$.
The difference between the two terms, related by the required Hermiticity property of the sampled matrix, can then be used as a metric of the quality of off-diagonal sampling of 
the RDM, before it is averaged to become strictly Hermitian at the end of the calculation (numerically equivalent to including both contributions at the accumulation stage).

These stochastic reduced density matrices do not satisfy the trace relation, given (for the 2-RDM) by
 \begin{equation}
     \sum_{pq} \Gamma_{pq,pq} = \frac{N(N-1)}{2} \hspace{1cm} p > q 	,\label{eqn:RDMTrace}
\end{equation}
where $N$ is the number of active electrons. This is because the $n_{\bfi}$ values are not normalized determinant coefficients.
This is more difficult to account for directly, as normalising the full wavefunction requires average sign values for all determinants, not just those instantaneously occupied.
Instead, the 2-RDM is normalised directly at the end of the simulation to enforce the trace relationship in \rff{eqn:RDMTrace}, which we are free to do as this normalization is a
freedom in the sampling of the wavefunction. Further N-representability constraints would be hard to apply, since these
generally correspond either to structural relationships between higher-ranked RDMs (which we do not have access to in this scheme) and lower ranked ones, or constraints on
the spectrum of the reduced density matrices which would be difficult to constrain at the sampling stage\cite{Coleman1963,Percus1964,ZhaoBraamsFukudaOvertonPercus2004}. 
Furthermore, we believe in this scheme that the RDMs obtained should 
as faithfully as possible reflect the wavefunction which is sampled, rather than imposing additional constraints, which would result in derived properties not directly corresponding 
to expectation values of the sampled wavefunction. It should be stressed that if the walker distribution is sampling the correct wavefunction, then all N-representability 
conditions will be exactly satisfied upon appropriate averaging in imaginary time.

The technique outlined above allows for an accumulation of an approximate RDM, alongside the stochastic dynamic of the FCIQMC wavefunction. However, there are a couple of
subtleties which need to be considered. The first concerns the fact that off-diagonal contributions to the RDM arising from determinant pair $\{D_{\textbf{i}}$, $D_{\textbf{j}}\}$ 
are only included in the event of a successful spawning event from one to the other. The probability of this occurring is proportional to the Hamiltonian matrix element
connecting the two determinants (in the case of a stochastic spawning event), which means that determinant pairs whose spawning events are discretised and connected 
via large Hamiltonian matrix elements will be sampled more frequently than weakly 
coupled determinants. This can result in sampling difficulties if two significantly weighted determinants are {\em not} in the non-integer walker space, and are connected via 
very small matrix elements, since a contribution to the RDM may be poorly sampled, or even entirely omitted. It should be stressed that continuous spawning connections
are not sampled proportionally to the Hamiltonian matrix element, and so are not affected.

A specific example where issues can arise concerns the connections to single excitations of the Hartree--Fock determinant ($D_{\textrm{HF}}$) in a canonical 
basis, whose connecting matrix 
elements are therefore numerically zero by Brillouin's theorem. To account for this, connections between the reference (generally Hartree--Fock) 
determinant and its excitations (both single
and double excitations) are taken into account explicitly. This is simple to do since the weight on the reference determinant is known to all parallel processes, as it is used for 
the calculation of the projected energy estimate, $E_P$. Since the contribution from these pairs of determinants is performed exactly (which are likely 
to be the main contribution, especially in relatively weakly correlated systems), their contributions are not dependent on a successful spawning event between the two. 

However, it is still possible that other very small Hamiltonian matrix elements connect substantially weighted determinants, and a sampling error could then arise. 
This was investigated by including additional spawning events between determinants. These events would spawn walkers between determinants proportionally to a function of the 
{\em inverse} of the Hamiltonian matrix element between them, or a specific lower-bound cutoff. These spawned walkers would not adjust the walker amplitude on 
determinants they were spawned to, or affect the simulation trajectory, but rather would simply confer information about the respective determinant parent and its 
amplitude (with appropriately adjusted $p_c$) in order to include the contribution to the off-diagonal elements of the RDMs when the Hamiltonian matrix element 
between the two determinants was small. This was found to provide negligible improvement across a range of systems as this sampling error was small, and therefore we consider 
this unnecessary and do not pursue it further here.

Despite this, the quality of the sampled density matrices remains poor, and convergence to the exact density matrix and derived properties with increasing walker number
is slower than for properties such as the projected energy of the system (see results in Ref.~\onlinecite{Booth2012}). This is due to a remaining significant error beyond simply a
manifestation of insufficient sampling time or initiator error of the wavefunction. This remaining bias can be summarized by the observation that the appropriate 
contribution to the density
matrix elements, $\bra n_{\textbf{i}}(\beta) \ket_{\beta} \bra n_{\textbf{j}}(\beta) \ket_{\beta}$, is in fact approximated by $\bra n_{\textbf{i}}(\beta) n_{\textbf{j}}(\beta) \ket_{\beta}$. This
ignores the non-zero (co)variance between the two amplitudes, which (especially for diagonal contributions) can be significant, and results in a biased sampling, even if the 
averaged walker amplitudes themselves are unbiased. The root cause is therefore the fact that the two amplitudes which contribute to the 
density matrix element are correlated, with the dominant error unsurprisingly coming from the diagonal elements where the instantaneous amplitudes are perfectly correlated, and
where the error is of a single sign, removing the possibility of error cancellation in the neglected (co)variance. 
%Essentially, for the diagonal density matrix elements, the random errors become systematic errors in this approach.

Computing appropriately averaged values of the determinant coefficients over the simulation, $\bra {n}_{\textbf{i}}(\beta) \ket_{\beta}$, which would remove the instantaneous
correlation between the walker populations, has prohibitive
memory demands as it involves storage of determinants that are occupied at \emph{any} stage of the simulation, even if their instantaneous population is zero. A simple
approach to reduce the bias is to average the walker populations during the period over which the determinant is occupied and use this as an approximate value for the
$\bra {n}_{\textbf{i}}(\beta) \ket_{\beta}$ determinant amplitude for the density matrix contribution (as was performed for Ref.~\cite{Booth2012} and the `biased' 
comparison results in section~\ref{sec:newRDMmethod_algo}). However, since this contribution is not averaged over periods while the
determinant is unoccupied, the density matrix is still a biased expectation value of the sampled wavefunction. Numerical results suggest that although this approach will tend
to the exact density matrix and associated properties in the large walker limit, the convergence to the exact result requires far more walkers compared to the projected
energy estimate.

In a recent article\cite{Blunt2014} Blunt \emph{et al.} introduced the `density matrix quantum Monte Carlo' (DMQMC) method. In this, the {\em full} $N$-electron density matrix
is sampled stochastically. From this quantity, lower-rank reduced density matrices can be obtained via `tracing out' of additional degrees of freedom. However, this approach
contrasts with the approach taken here, where only the wavefunction is sampled, and we seek to directly find the reduced density matrix which corresponds to this sampled
wavefunction. However, we draw inspiration from the DMQMC approach, since they considered the use of a replica trick to overcome a similar difficulty relating to the evaluation
of a quadratic function of a stochastically sampled distribution (this time of the density matrix rather than wavefunction) -- the Renyi $S_2$ entanglement 
entropy. In this, two uncorrelated walker samples were propagated 
independently, thus rigorously removing any covariance between the sampled distributions. They further suggest that this approach might be useful for sampling 
expectation values, such as in Eq.~\ref{eqn:PureEx}, directly within FCIQMC. This approach was subsequently used in Ref.~\onlinecite{kpfciqmc}, in order to 
compute {\em dynamical} expectation values, including the Green's functions and arbitrary spectra. In the rest of this article we take up this idea to calculate RDMs 
within FCIQMC and demonstrate that an unbiased sampling can indeed be achieved, dramatically improving the convergence over the biased approach.
%In the next section, we detail a new approach to rigorously remove this sampling bias by allowing for {\em two} uncorrelated walker samples (replicas), which removes
%the covariance between the sampled wavefunctions of the `bra' and `ket', and dramatically 
%improves the convergence of the density matrices, pure expectation values and variational energy estimator\cite{Blunt2014,kpfciqmc}. The improvement 
%allows for the convergence with number of walkers of properties which is comparable to that of the projected energy estimate, with the initiator
%error in the wavefunction itself as the remaining systematic error.

\section{Unbiased FCIQMC Density Matrix Replica Sampling} \label{sec:newRDMmethod_algo}

An unbiased RDM sampling can be achieved via two independent walker populations (replicas), whose instantaneous weights on a determinant $D_{\bfi}$ will be denoted
$n_{\textbf{i}}^{(1)}$ and $n_{\textbf{i}}^{(2)}$ for the first and second walker replicas respectively. In this implementation, the replicas are 
propagated with independent random number strings, and each determinant with a walker weight in either of the two replicas is stored in the data 
structures outlined in Ref.~\onlinecite{BSA2014}, in the same fashion as handling complex walker amplitudes in Ref.~\onlinecite{BGKA2013}. 
This allows for the weight on a given determinant over both replicas to be extracted without any additional computational overhead.

Spawning, death and annihilation events are then applied to each population independently and two sets of 
statistics gathered, including measures of $E_P$ and $E_S$, with $E_S^{(1)}$ and $E_S^{(2)}$ used in the respective death probabilities for the replicas.
The two simulations are therefore completely uncorrelated, apart from sharing the same initial conditions, from which they quickly decorrelate after equilibration. Although this 
effectively doubles the computational resources required per iteration, there are two independent estimates of all quantities of interest, and so by using this additional information, 
the random errors in these quantities are reduced by a factor of $\sqrt{2}$. This results in the calculation requiring only half the time. Computational 
overheads therefore only effectively result from increased memory requirements from the additional replica population.

Uncorrelated and unbiased contributions to elements of the reduced density matrices can then be found by ensuring that any product of determinant populations is computed from
the amplitudes over both replicas. Contributions to the diagonal elements of the RDMs can be gathered each iteration from all instantaneously occupied determinants, by accumulation
of $n_{\textbf{i}}^{(1)}(\beta) \hspace{2pt} n_{\textbf{i}}^{(2)}(\beta)$ to the $\frac{N(N-1)}{2}$ relevant diagonal elements. Off-diagonal elements are accumulated through
successful spawning events, as detailed in section~\ref{sec:SamplingRDMs}, ensuring that the two determinant weights $n_{\textbf{i}}$ and $n_{\textbf{j}}$ come 
from different populations to ensure an unbiased sampling. However, there are two possibilities with this.

From consideration of a determinant transition from replica 1, $D_{\bfi} \to D_{\bfj}$, which occurs with probability $p_c(\bfj|\bfi)^{(1)}$, two possible contributions could be considered;
\begin{align}
 \textrm{Forwards} \hspace{7mm}    &\frac{n_{\textbf{i}}^{(1)}n_{\textbf{j}}^{(2)}}{p_c(\textbf{j}|\textbf{i})^{(1)}}		\label{eqn:Forwards}		\\
 \textrm{Backwards} \hspace{7mm} &\frac{n_{\textbf{i}}^{(2)}n_{\textbf{j}}^{(1)}}{p_c(\textbf{j}|\textbf{i})^{(1)}}	.	\label{eqn:Backwards}
\end{align}
It is important that only the {\em `forwards'} contribution is considered for the spawning events from replica 1 (Eq.~\ref{eqn:Forwards}), since including the latter `backwards' 
contribution (Eq.~\ref{eqn:Backwards}) introduces a bias into the sampling. This can be rationalized from consideration of an iteration where $n_{\bfi}^{(1)}=0$ and $n_{\bfi}^{(2)} \neq 0$.
In this case, there would be non-zero $n_{\textbf{i}}^{(2)}n_{\textbf{j}}^{(1)}$ contributions which would never be sampled, because there are no walkers on site $\bfi$ in the first replica to spawn from.
Instead the `backwards' contribution should be sampled only from transitions within replica 2, with the denominator 
changed to $p_c(\textbf{j}|\textbf{i})^{(2)}$, and appropriately averaged with the forwards contribution. However, for the results presented in this paper, the 
contributions from only one replica were considered, indicating 
that a further reduction of the error bars in the density matrix elements by a factor of $\sqrt{2}$ is trivially possible without additional cost.

% - PLOT - comparison of the averaged method with with SR biased method (C_2)
% It would be really nice to get error bars on the RDM values, but I do not want to rerun all of these calculations... anyone want to volunteer?
\begin{figure*}[htp]
\centering
\includegraphics[width=0.95\textwidth]{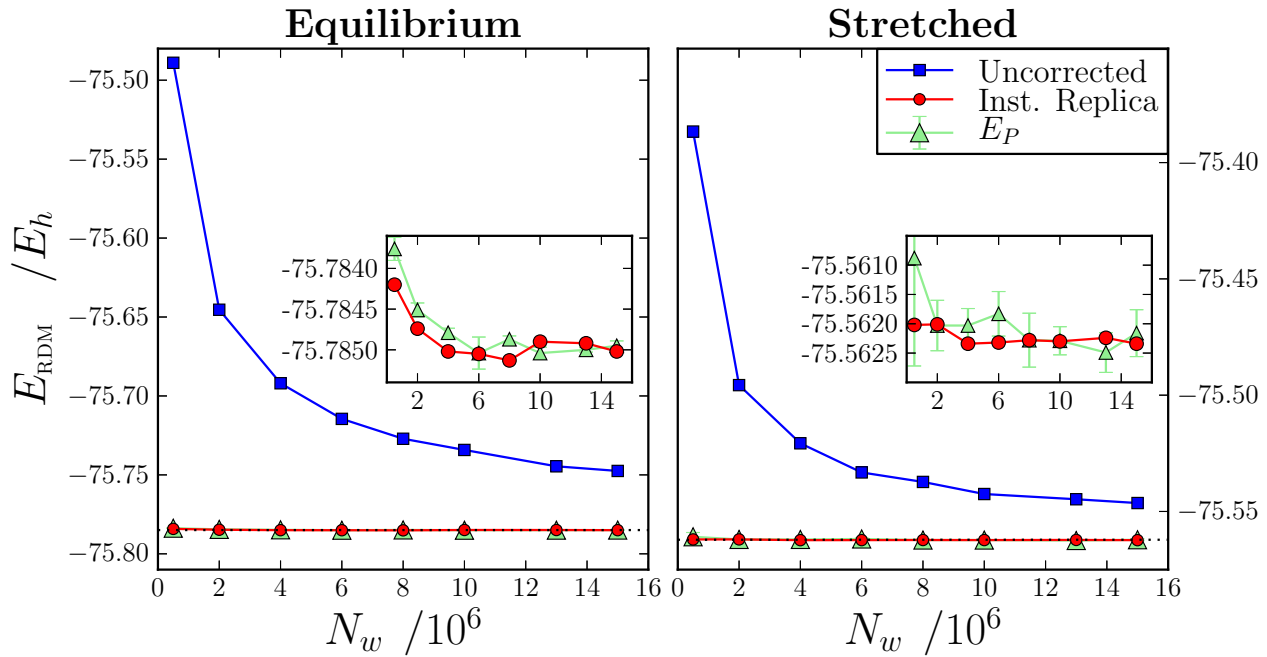}
\caption{
    Comparison of the convergence with number of walkers ($N_w$) for the energy estimate calculated from the sampled density matrices ($E_{\textrm{RDM}}$ as given 
    in Eq.~\ref{eqn:ERDM}), for C$_2$ in a cc-pVTZ basis. The Hilbert space size is $\mathcal{O}[10^{10}]$ many-body functions. `Uncorrected' refers to the biased sampling detailed in section~\ref{sec:SamplingRDMs}, while the 
    expectation values from the unbiased replica-sampled RDMs of section~\ref{sec:newRDMmethod_algo} are denoted `Inst. Replica'. Also for comparison is the 
    projected energy estimate, $E_P$. This system is identical to the one studied in Ref~\onlinecite{Booth2012}, which used the biased sampling algorithm, 
    athough all values have been newly generated here.
    Uncorrected RDMs were calculated with the integer i-FCIQMC algorithm (consistent with Ref.~\onlinecite{Booth2012}), whilst
    replica-sampled RDMs use the non-integer walker weights as detailed in section~\ref{sec:reals}, with $\chi=4$, $N_{\textrm{occ}}=1$, to represent the best quality 
    RDM available with the techniques presented in this paper. Errorbars are not included for $E_{\textrm{RDM}}$ estimates, since only one calculation was performed.
    The use of non-integer walker weights does not generally remove the systematic error, and so the improvement in $E_{\textrm{RDM}}$ between the two sampling
    techniques can be considered a result of the use of the unbiased replica sampling. Remaining systematic error at low walker numbers can be attributed to the 
    convergence of the initiator approximation. This manifests more strongly in the equilibrium results than the multiconfigurational stretched case.
    }\label{fig:C2_F12paper_new}
\end{figure*}

Finally, as described in section~\ref{sec:SamplingRDMs}, the off-diagonal contributions that involve direct connections to the reference 
determinant (generally $D_{\textrm{HF}}$) can be included
explicitly, with the knowledge of $n_{\textrm{HF}}$ for both replicas known to all processes each iteration. This will ensure the best quality sampling for these generally important
contributions, and avoid any sampling issues arising from Brillouins theorem. Again, the sampled density matrices are normalized to fulfil the appropriate trace relations, as well
as made Hermitian at the end of the simulation (which is only exactly achieved without this in the limit of long sampling time), which ensures 
that $\Gamma_{pq,rs} = \Gamma_{rs,pq}^{\ast}$ for the 2-RDM.

A simple comparison between the original, biased approach, and the replica sampling of the density matrices in Fig.~\ref{fig:C2_F12paper_new} shows the 
striking improvement in the quality of the properties. It can be seen that the convergence of the quasi-variational energy expectation value, $E_{\textrm{RDM}}$, is now 
comparable to that of the non-variational estimate, $E_P$ (or faster in some instances). The $E_{\textrm{RDM}}$ estimate is denoted `quasi-variational', since it is only 
rigorously variational in the long sampling
limit. For short sampling times, it is possible that the replicas have sampled different wavefunctions within their stochastic errors, and therefore the $N$-representability conditions
of the RDMs will not be strictly satisfied. Non-variational energies are therefore possible within the random errors of the sampling.

In the limit of sufficient sampling of imaginary time, but where the number of walkers is not sufficient to entirely converge the initiator error of the FCIQMC 
method, $E_{\textrm{RDM}}$ will be variational, but is not in general the same as $E_P$. These two energy estimates will only rigorously be the same where the sampled 
wavefunction represents an eigenstate. Therefore, this
quasi-variational estimate of the energy may have a different convergence profile with $N_w$ compared to the standard energy estimates. This might provide a clearer measure of
decay of the initiator error, as well as potentially being more amenable to variational extrapolation of this energy to the infinite walker limit to remove the effects of the initiator error.
This will be investigated in future work.

\begin{figure*}[htp]
\centering
\includegraphics[width=0.95\textwidth]{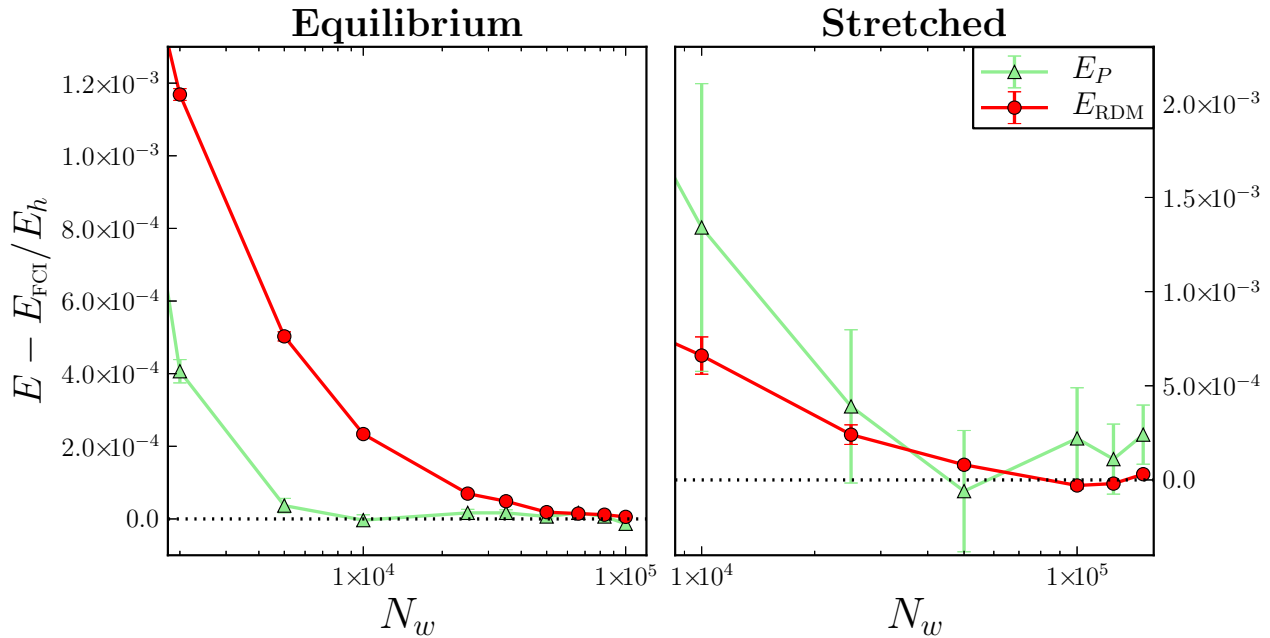}
\caption{
    Convergence of $E_{\textrm{RDM}}$ and $E_P$ with $N_w$ for equilibrium (1.094\AA) and stretched (4.2a$_0$) $\textrm{N}_2$ in a cc-pVDZ basis.
    $E_{\textrm{RDM}}$ was calculated with the instantaneous replica sampling method, using non-integer coefficients. $\tau=10^{-3}$a.u.$^{-1}$ and 8 electrons were frozen. The Hilbert space size is $\mathcal{O}[10^5]$ many-body functions.
    Convergence rate is seen to be substantially improved (comparable to or faster than the projected energy estimate) for more multiconfigurational systems.
}\label{fig:ChoosingNw}
\end{figure*}

The convergence to this large walker limit of the $E_{\textrm{RDM}}$ expression can be assessed via comparison to FCI results, as shown in Fig.~\ref{fig:ChoosingNw}, where a
smaller system than that of Fig.~\ref{fig:C2_F12paper_new} allows the exact 2-RDM to be calculated. This indicates that the $E_{\textrm{RDM}}$ expression appears to provide a
more rapid convergence to the large walker limit in the case of more multiconfigurational wavefunctions, simulated by the stretching of the triply-bonded nitrogen 
molecule to 4.2~a$_0$. This is unsurprising, since the projected energy expression used a trial wavefunction which was simply the Hartree--Fock determinant, whose weight
in the FCI expansion diminishes rapidly as the bond is stretched\cite{ChanKallayGauss2004}. The use of a multiconfigurational trial wavefunction would 
improve the projected energy
estimate in these cases. However, it is clear that in these cases, $E_{\textrm{RDM}}$ provides a good estimate of the energy, with a far larger number of spawning events in 
the space directly contributing to the energy (all to a determinant with a non-zero population in the other replica), rather than just the distribution of a small fraction 
of walkers on the reference determinant and its direct excitations.

\begin{figure*}[htp]
\centering
\includegraphics[width=0.95\textwidth]{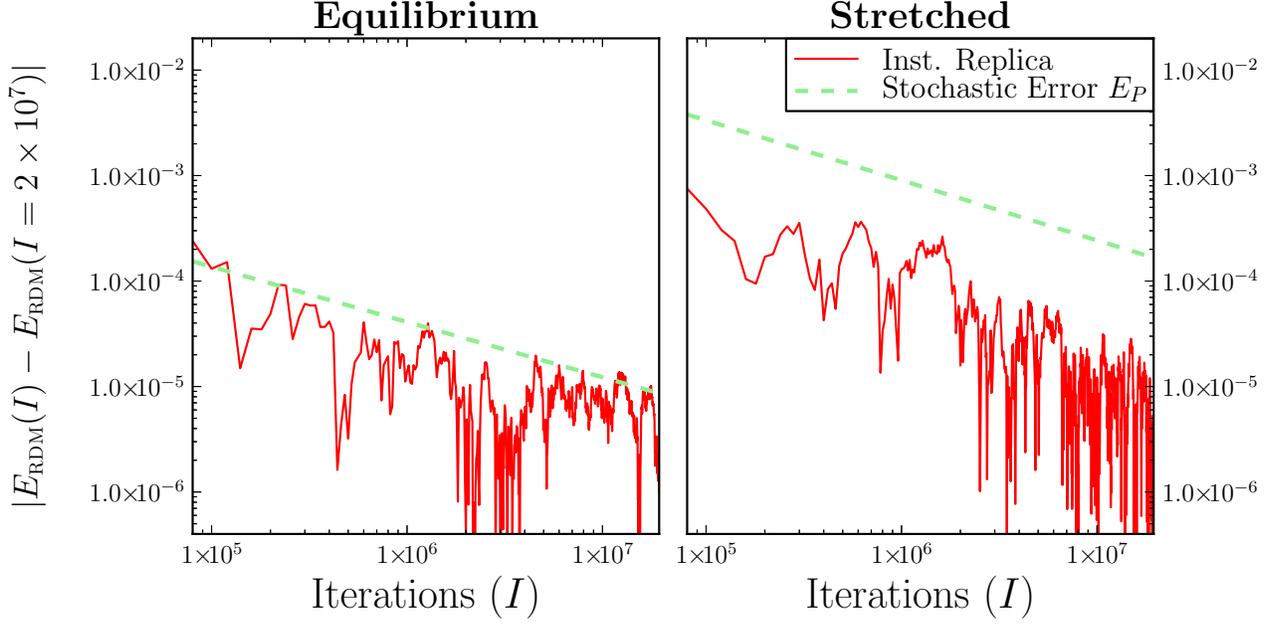}
\caption{
    Convergence of errors in $E_{\textrm{RDM}}$ and $E_P$ with Monte Carlo iterations for equilibrium and stretched $\textrm{N}_2$.
    $E_{\textrm{RDM}}$ values are calculated with the replica sampling method and non-integer coefficients.
    The values at each accumulation time are then plotted relative to the value at $I=2 \times 10^7$, to give an estimate of the random sampling error 
    in $E_{\textrm{RDM}}$ with iterations $I$.
    The dashed green line represents the random error in $E_P$ calculated over the same number of iterations and simulation setup.
    This was calculated by conducting a blocking analysis on data sets of differing length, and fitting to these data points\cite{Flyvbjerg1989}.
    The plotted fit lines for the $E_P$ estimate have the form $\mathcal{O}[I^{-0.52}]$ and $\mathcal{O}[I^{-0.57}]$, consistent with the expected $\mathcal{O}[I^{-0.5}]$ scaling of 
    stochastic errors in Monte Carlo methods.
    It can be seen that the error in the $E_{\textrm{RDM}}$ estimate from the accumulated RDMs is (at all iteration values) less than the corresponding $E_P$ energy,
    while following the same $\mathcal{O}[I^{-0.5}]$ rate of convergence. This
    indicates that the random errors in the $E_{\textrm{RDM}}$ energy are equivalent, or smaller, than the corresponding $E_P$ value for a given number of iterations, 
    with the stretched case showing a greater improvement over $E_P$.
}\label{fig:simulationtime}
\end{figure*}

This improvement of the $E_{\textrm{RDM}}$ energy estimator over the $E_P$ value is again borne out for multiconfigurational problems when considering not the number
of walkers required, but the sampling time required for the random errors associated with the energy estimate to fall to an acceptable value. This can be observed 
in Fig.~\ref{fig:simulationtime}, where the convergence of the energy for a given number of iterations is in general faster than the corresponding projected energy estimate.
The convergence of other properties derived from the sampled density matrices will be investigated in section~\ref{sec:Properties}, after a consideration of the computational
costs associated with sampling of the density matrices.

% - To save time and effort, we may want to average -- explain how this is done in an unbiased way.
% - Mention cost
% - Mention other stuff (multireference, etc).
\subsection{Computational Efficiency Considerations}

The replica-sampled RDM method achieves an unbiased sampling of the 2-RDM by adding in instantaneous contributions 
$n_{\textbf{i}}^{(1)/(2)} \times n_{\textbf{j}}^{(2)/(1)}$ throughout the simulation. Although there is an associated computational cost for running another independent 
replica of the walker population, this cost is largely offset by the improvement in the statistics that result compared to running twice as long. Therefore, the
dominant additional overhead for propagation of the additional replica comes from the associated memory requirements. The accumulation of the 2-RDM also has
associated costs, since the memory required is of size $\mathcal{O}[M^4]$. Since this is of the same size as the storage of the two-electron integrals required for the
calculation and also stored in memory, it is currently not a bottleneck. However, eventually these memory costs may be required to be distributed among the 
computational nodes\cite{BSA2014}.

The dominant computational cost in sampling the density matrix comes from the accumulation of the diagonal matrix elements, as each occupied 
determinant requires an $\mathcal{O}[N^2]$ operation to identify and update all the necessary 2-RDM elements each iteration.
Therefore, any approach that reduces the frequency at which these contributions need to be added in will reduce the computational cost 
associated with accumulating the RDM, with increasing returns at large $N$.
This can be achieved with an approach whereby we store and calculate a running average of a determinant's weight within a 
defined block of iterations, and only add in its diagonal RDM contributions (and connections to reference determinant, if applicable) once at the end of 
these blocks, weighted by the number of iterations in that block.
In so doing, contributions to the diagonal elements and $D_{\textrm{HF}}$ connections can be summed in much less frequently without any loss of information.

However, the implementation of such an averaging technique requires careful thought to ensure that no systematic bias is introduced into the RDM contributions, 
whilst making the best use of the available data in the simulation. This dictates that `blocks' of iterations over which the weight on a determinant is averaged should start
whenever a determinant has a change in occupation status (i.e. becomes newly unoccupied or newly occupied) {\em in either replica population}. This ensures that the 
only blocks with a non-zero contribution to the RDM exactly correspond with the set of iterations that would have produced a non-zero contribution when adding 
in snapshots instantaneously every iteration.

This averaging approach retains the unbiased nature of the RDM sampling, but has a significant advantage in terms of simulation time, as contributions to the 
diagonal elements and $D_{\textrm{HF}}$ connections are only required at block boundaries (and for the diagonal contribution, only when both replicas have been 
occupied in the block), rather than every iteration.
Therefore, for a given determinant, its contributions to the RDM diagonal elements, which are added in only at the end of each averaging block, now take the form
\begin{equation}
    \bra n_{\textbf{i}}^{(1)}\ket {}_a \hspace{2pt} \bra n_{\textbf{i}}^{(2)}\ket_a I(D_{\textbf{i}})_a
\end{equation}
where $\bra n_{\textbf{i}}^{(1)}\ket {}_a$ is the average weight on $D_{\textbf{i}}$ in population 1 within averaging block $a$, and  $I(D_{\textbf{i}})_a$ is the number of iterations within that averaging block. This removes what rapidly becomes the dominant cost as the number of electrons in the system increases.
Similarly, if the given determinant is a single or double excitation of $D_{\textrm{HF}}$, the off-diagonal contributions from this connection are added in at the same time as the diagonal contributions, taking the form:
\begin{equation}
    \bra n_{\textrm{HF}}^{(1)} \ket \hspace{2pt} \bra n_{\textbf{i}}^{(2)} \ket_a I(D_{\textbf{i}})_a
\end{equation}
where $\bra n_{\textrm{HF}}^{(1)} \ket$ is the measure of the average population on $D_{\textrm{HF}}$ up to the iteration where the contribution is included.
%IS THIS BIASED? WHY SHOULDN'T n_HF HAVE TO BE REAVERAGED FOR EACH BLOCK TOO?

Although this averaging approach can also be shown to achieve an unbiased sampling of the diagonal elements and the explicitly included $D_{\textrm{HF}}$ connections, 
the presence of serial correlation in determinant amplitudes creates a problem for the remaining stochastically sampled off-diagonal contributions if these averaged 
coefficients are used.
%This can be justified first by recalling that a new average $\bra n_{\textbf{i}}^{(1)} \ket_{\tau}$ is begun when either $n_{\textbf{i}}^{(1)}$ or $n_{\textbf{i}}^{(2)}$ change their occupation status (i.e. from occupied to unoccupied, or vice versa).
%Therefore in a significant portion of these cases, $n_{\textbf{i}}^{(1)}$ has just been reoccupied.
%As the determinant populations are serially correlated, the instantaneous values of $n_{\textbf{i}}^{(1)}$ at the start of the averaging block will be close to zero, meaning that the cumulative average $|\bra n_{\textbf{i}}^{(1)} \ket_{\tau}|$ will almost invariably converge from below in these cases, leaving a strong residual bias.
%%As the occupation status of $n_{\textbf{i}}^{(2)}$ is completely uncorrelated from $n_{\textbf{i}}^{(1)}$, other instances where $\bra n_{\textbf{i}}^{(1)} \ket_{\tau}$ is rezeroed will lead to roughly even convergence from above and below, leaving a strong residual bias in the value of $\bra n_{\textbf{i}}^{(1)} \ket_{\tau}$.
%This effect does not cause trouble for the contributions to diagonal elements and $D_{\textrm{HF}}$ connections as the full occupation period is specifically accounted for, but does cause a bias for the stochastically sampled off-diagonal elements, as the value of $\bra n_{\textbf{i}}^{(1)} \ket_{\tau}$ at any given point is assumed to be representative of the full mean population $\overline{ n_{\textbf{i}}^{(1)}}$.
Therefore, off-diagonal elements must continue to be accumulated using instantaneous determinant weights as detailed previously. However, this is not a particular burden
since for the off-diagonal elements there is only one contribution (for sampled double excitations), or $\mathcal{O}[N]$ (if a single excitation is sampled).

\section{Molecular Properties}	\label{sec:Properties}

% - PLOT - N2 eqm - properties convergence 158 but without cutoff and inst.
% - PLOT - N2 str - properties convergence 161 but without cutoff and inst
% - PLOT - N2 eqm and str - tau convergence of energy - either 136 or 146?
% - PLOT - from p145

It is important not to simply consider the density matrix energy estimator when benchmarking the quality of the RDMs, since the action of contracting the density matrices with
the Hamiltonian will place emphasis on parts of the RDM matrix, while being relatively insensitive to others. Indeed, the algorithm dictates that it preferentially samples
those elements whose contribution to the variational energy estimator is largest. In addition, the primary motivation for the sampling of the density
matrices is for the computation of unbiased molecular properties derived from the density matrices, whose value could not be computed in an unbiased fashion from a
projected estimator.

\begin{figure}[ht]
\centering
\includegraphics[width=0.95\columnwidth]{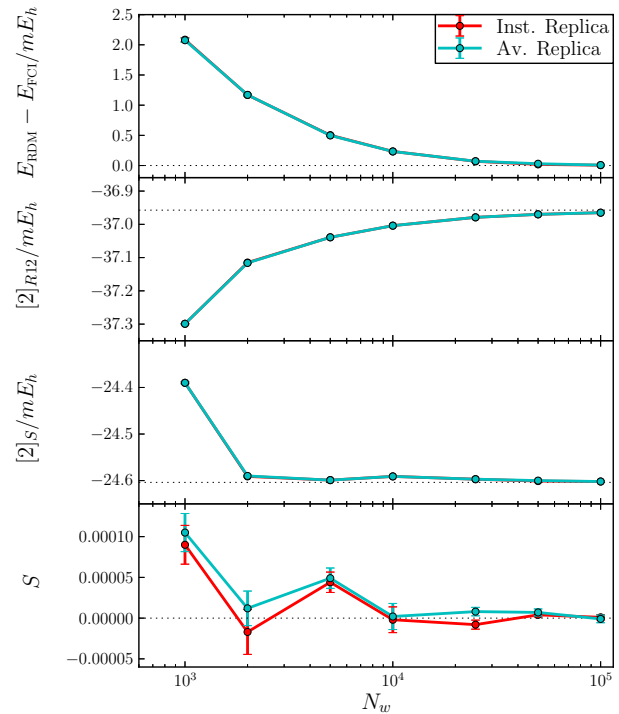}
\caption{
    Comparing instantaneous and averaged diagonal RDM replica sampling methods for convergence of selected properties with number of walkers, $N_w$, 
    for equilibrium (1.094\AA) N$_2$ cc-pVDZ. Both approaches are unbiased, and are shown to be equivalent within their random errorbars (most are entirely overlaid). 
    Plotted is $E_{\textrm{RDM}}$, as well as $[2]_{R12}$ and $[2]_S$ explicit correlation energy corrections (see 
    Refs.~\onlinecite{Torheyden:JCP131-171103,Kong:JCP133-174126}) and the value of the spin quantum
    number, $S$, of the sampled wavefunction\cite{ZhaoBraamsFukudaOvertonPercus2004}. 
    These errorbars are calculated from five independent calculations, but are often too small to be seen.
    Dotted lines give the exact FCI value for each quantity, while stochastic RDMs were accumulated over $\sim 5\times 10^6$ iterations.
    8 electrons are frozen, with $\tau=10^{-3}$a.u.$^{-1}$}
    \label{fig:properties_eqm_cheaper}
\end{figure}

\begin{figure}[!ht]
\centering
    \includegraphics[width=0.95\columnwidth]{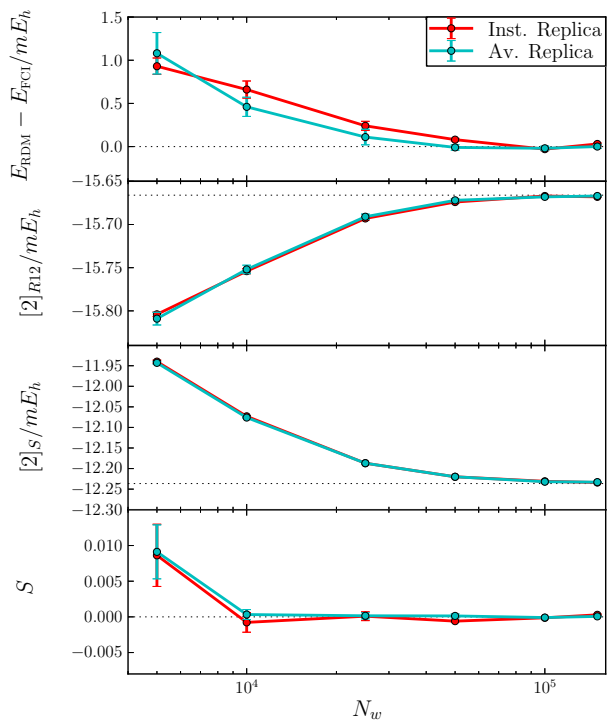}
\caption{This plot details equivalent results to Fig.~\ref{fig:properties_eqm_cheaper}, but for the more multiconfigurational, stretched 
N$_2$ molecule ($4.2 a_0$). Note the reduction in the error in $E_{\textrm{RDM}}$ as the bond is stretched. This contrasts in general with $E_P$,
which degrades as the wavefunctions become less compact and more difficult to sample in these multiconfigurational cases\cite{Thomas2014}.}
    \label{fig:properties_str_cheaper}
\end{figure}

Figures~\ref{fig:properties_eqm_cheaper} and \ref{fig:properties_str_cheaper} show various properties derived from the one- and two-body sampled density 
matrices, compared to exact results from density matrices computed from a FCI diagonalization of the system. Both an equilibrium (relatively single reference), 
and stretched triple bond (strongly multiconfigurational) are shown, to highlight any potential differences arising from the sampling of the RDMs within different wavefunction characteristics. Also, both the instantaneous replica RDM sampling, as well as the cheaper but still unbiased sampling of averaged determinantal 
weights for diagonal RDM elements are shown.
These are all within random errorbars of each other, confirming the equivalence of these two approaches, and the fact that the averaging of determinant weights should
always be performed.

The properties sampled in these results include the spin quantum number of the sampled wavefunction\cite{ZhaoBraamsFukudaOvertonPercus2004}, 
and two differing corrections to the wavefunction ameliorating the 
basis set incompleteness in the representation, proposed by Valeev {\em et al.}\cite{Torheyden:JCP131-171103,Kong:JCP135-214105,Kong:JCP133-174126}. 
The first ($[2]_{R12}$) approximately corrects for the incompleteness in the two-electron
form of the correlation hole around each coalescence point\cite{Torheyden:JCP131-171103}, while $[2]_{S}$ corrects for the incompleteness relating to the 
single-particle orbital representation\cite{Kong:JCP135-214105}. Although the $[2]_{S}$ correction was originally proposed in the context of accelerating convergence
with respect to the external space within CASSCF calculations, in this context it is used without orbital optimization, with the entire orbital space considered as occupied. This
can cause convergence difficulties and intruder states as $H^{(0)}$ does not necessarily separate between the orbital and CABS spaces. However, whether this quantity is a 
robust correction for single-particle
basis incompleteness without combining it with orbital optimization is separate to the use of it as a metric for analysis of the convergence of the parent density matrix, for which it is primarily used in this study.

These quantities,
as well as other explicit correlation corrections, have been applied to FCIQMC wavefunctions previously\cite{Booth2012,Sharma2014}. It was found that these two properties 
were far more 
sensitive than $E_{\textrm{RDM}}$ on the quality of the sampled off-diagonal 2-RDM elements, and so provided a complementary test of the sampling quality. Nevertheless,
the errors in these quantities is an order of magnitude smaller than $E_{\textrm{RDM}}$, and converges at least as rapidly to the exact result. The error in these quantities is
expected to solely result from initiator error in the sampled wavefunction, which decreases as observed as the number of walkers increases, as all other sources 
of systematic error from the density matrices have been removed\cite{footnote2,ThomPopBias2014}.
%It should be noted that there is also potential for a small bias in the wavefunction sampling (which
%is also systematically improvable with increasing walker number) due to the fluctuation in the $E_S$ parameters and consequent changes in walker population. However, this
%error is generally only $\mathcal{O}[10^{-5}]E_h$, and so can be considered negligible for these system\cite{ThomPopBias2014}.}
The random error in the calculations is also very small, having been estimated from independent calculations.

\section{Summary and Conclusions}

In summary, we propose the replica-sampled RDM method for computing one- and two-body reduced density matrices from a stochastic FCIQMC sampling of the
wavefunction. The replica sampling allows for the elimination of systematic errors due to the quadratic dependence of the density matrix elements on the sampled wavefunction,
from which a previous implementation suffered badly. The resultant reduced density matrices have no systematic error besides the initiator error present in the inherent sampling
of the wavefunction itself, which can be systematically improved by increasing overall walker number. The accumulation of the RDMs runs alongside the sampling of the wavefunction,
making use of the two-body operator moves of FCIQMC, therefore adding only modest additional computational overheads for the calculation of the density matrices. 

Averaged determinant weights are used for the contributions to the RDM diagonal elements and $D_{\textrm{HF}}$ connections, therefore avoiding a potentially 
costly $\mathcal{O}[N^2]$ step for each occupied determinant each iteration, without additional errors being incurred. However, instantaneous 
weights are used for the stochastic sampling of the remaining off-diagonal contributions, retaining the unbiased qualities of the original all-instantaneous 
replica sampling approach. The properties derived from these density matrices are shown to converge with increasing walker number to exact results 
at a comparable rate to the standard projected energy estimators, as the initiator error is reduced.

This is also true for the quasi-variational energy estimator derived from the density matrices (strictly variational in the large imaginary-time limit).
This estimate is found to be an accurate alternative to the projected estimate, especially powerful in strongly correlated, multiconfigurational wavefunctions, where
the absence of a large weighted wavefunction space to project onto can result in large random errors in this approach. Furthermore, the quasi-variational nature of
the value may lend itself to extrapolation techniques for the systematic initiator error. 

On the face of it, it would appear that the scaling of the replica-sampled reduced density matrix will be poor with respect to increasing system size, as it requires spawning events 
onto a determinant occupied in the other replica in order to accumulate off-diagonal contributions. However, it is precisely this type of event which is required for the annihilation
part of the algorithm, crucial for the suppression of the sign-problem in the space (aided by the initiator conditions). Because of this, we do not anticipate the scaling to be
particularly worse than that of the fundamental ability of the algorithm to resolve the correct wavefunction. This precise scaling is expected to be dependent on the expectation
values being probed, as well as the system under investigation, and further studies are underway to observe the accuracy of the properties as the system sizes increase.

Finally, it should be noted that it is straightforward to combine this approach with
more recent developments and extensions of FCIQMC, such as the semi-stochastic sampling of the 
space\cite{PetruzieloHolmesChanglaniNightingaleUmrigar2012}, extraction of excited states\cite{BoothChan2012,TennoMSQMC}, and additional symmetries such as
time-reversal symmetry and (angular) momentum symmetries which have been constrained in the walker dynamics\cite{BoothC2}. We now plan on turning our 
attention to larger systems, and an assessment of the accuracy of other properties, including nuclear forces and electrical moments of molecules.

\section{Acknowledgements}

GHB would like to gratefully acknowledge financial support from the Royal Society via a University Research Fellowship.
This work has been supported by EPSRC grant number EP/J003867/1. 

%\bibliography{RDMBib}

\begin{thebibliography}{60}
\bibitem [{foo({\natexlab{a}})}]{footnote1}%
  \BibitemOpen
  \href@noop {} {} \bibinfo {note} {We note that there has
  been recent work to also sample from a mixed distribution within an FCIQMC
  framework\cite{Clark2012}. However, as mentioned, this difference is not
  fundamental to the inherent bias or approach presented in this paper, which
  can also be employed within mixed distribution sampling of
  FCIQMC.}\BibitemShut {Stop}%
\bibitem [{foo({\natexlab{b}})}]{footnote2}%
  \BibitemOpen
  \href@noop {} {} \bibinfo {note} {It should be noted that
  there is also potential for a small bias in the wavefunction sampling (which
  is also systematically improvable with increasing walker number) due to the
  fluctuation in the $E_S$ parameters and consequent changes in walker
  population. However, this error is generally only $\mathcal{O}[10^{-5}]E_h$,
  and so can be considered negligible for these
  system\cite{ThomPopBias2014}.}\BibitemShut {Stop}%
\end{thebibliography}

\end{document}